\documentclass[fleqn,twoside]{article}
\usepackage{espcrc2}

\usepackage{epsfig}
\usepackage{graphics}
\usepackage{graphicx}

\newcommand{\AmS}{{\protect\the\textfont2
  A\kern-.1667em\lower.5ex\hbox{M}\kern-.125emS}}

\title{The CDF Calorimeter Upgrade for Run IIb}

\author{S.~Kuhlmann\address{\vspace*{-.1in}Argonne National Laboratory, Illinois 60439, USA},
   H.~Frisch\address{\vspace*{-.1in}University of Chicago, Chicago, Illinois 60637, USA},
   M.~Cordelli\address{\vspace*{-.1in}Laboratori Nazionali di Frascati, INFN, I-00044 Frascati, Italy},
   J.~Huston\address[MSU]{\vspace*{-.1in}Michigan State University, East Lansing, Michigan 48824, USA},
   R.~Miller\addressmark[MSU],
   S.~Lami\address{\vspace*{-.1in}INFN Pisa, I-56127 Pisa, Italy, and Rockefeller University, New York, New York 10021, USA}\thanks{Corresponding author. Tel.: +1-212-327-8832;
fax: +1-212-327-7786.  {\em E-mail address}: lami@fnal.gov
(S. Lami)},
   R.~Paoletti\address[Siena]{\vspace*{-.1in}INFN Pisa, I-56127 Pisa, and University of Siena, I-53100 Siena, Italy},
   N.~Turini\addressmark[Siena],
   M.~Iori\address{\vspace*{-.1in}University of Roma 1 and INFN, I-00185 Roma, Italy},
   D.~Toback\address{\vspace*{-.1in}Texas A\&M University, College Station, Texas 77843, USA},
   F.~Ukegawa\address{\vspace*{-.1in}University of Tsukuba, Tsukuba, Ibaraki 305, Japan}}

\begin{document}

\begin{abstract}
The physics program at the Fermilab Tevatron Collider will continue
to explore the high energy frontier of particle physics until the
commissioning of the LHC at CERN. The luminosity increase provided by the
Main Injector will require upgrades beyond those implemented for the
first stage (Run IIa) of the Tevatron's Run II physics program.
The upgrade of the CDF calorimetry includes: ~1)~ the replacement of the
slow gas detectors on the front face of the Central Calorimeter with a faster
scintillator version which has a better segmentation, and ~2)~ the addition
of timing information to both the Central and EndPlug Electromagnetic
Calorimeters to filter out cosmic ray and beam related backgrounds.
\vspace{-1pc}
\end{abstract}

\maketitle

\section{THE CENTRAL PRESHOWER AND CRACK DETECTORS}

The CDF {\it Central Preshower} (CPR) and {\it Central Crack} (CCR)
detectors will be replaced 
at the time the silicon detector is replaced for Run IIb.

\begin{figure}[hb]
\vspace*{-.335in}
\vspace*{.002in}
\centering{\epsfig{file=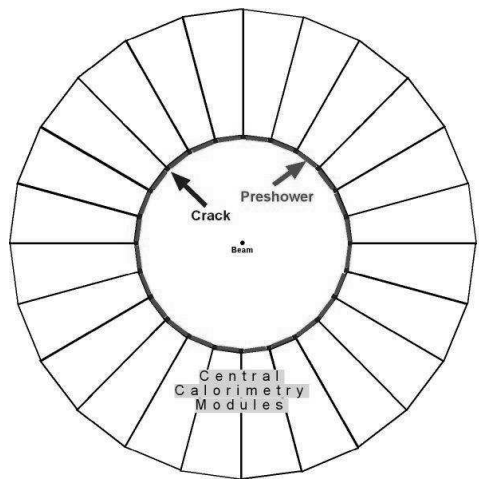,angle=0,width=15pc}}
\vspace*{-.36in}
\caption{The central calorimetry wedges and the
location of the preshower and crack detectors.}
\vspace*{-.08in}
\label{fig:calo_xsec}
\end{figure}

In 1992 the CDF Collaboration installed gas detectors
on the front face of the central calorimeter in order
to sample early showers and cover the $\phi$-cracks
between calorimeter wedges, as shown in Fig.~\ref{fig:calo_xsec}.
The CPR has been extensively used in electron identification (ID),
providing about a factor 2-3 more rejection of charged pions that pass
all other cuts. This extra rejection has been crucial
in soft electron ID for b-jet tagging, as was
shown in the first {\it top} evidence paper \cite{top_evid}.
The CPR has been used in several publications involving
photon ID. By using conversion rates, which are energy independent,
it extended the QCD measurement of direct photons by more
than 100 GeV in photon transverse momentum $P_T$ \cite{photons}.

The CCR, located after a 10 radiation length thick tungsten bar,
has been checked for large pulse heights in all the rare events CDF
has observed in Run I. The $\phi$-cracks cover about 8\% of the central
detector, and in events with multiple electromagnetic objects, the 
possibility of one object hitting the crack is quite high.

The present slow CPR and CCR gas
 detectors will suffer the
luminosity increase foreseen for Run IIb. In order to maintain the same
Run I capabilities, they will be replaced by scintillator counters
read out by {\it Wave-Length Shifting} (WLS) fibers.
The new CPR will also have a better segmentation and will be used
to improve the jet energy resolution by both correcting for energy loss in the
dead material in front of it and adding its information in jet algorithms
incorporating charged tracking.


\vspace*{-.05in}
\subsection{The Detector Design fron Run IIb}

\begin{figure}[htb]
\vspace*{-.36in}
\centering{\epsfig{file=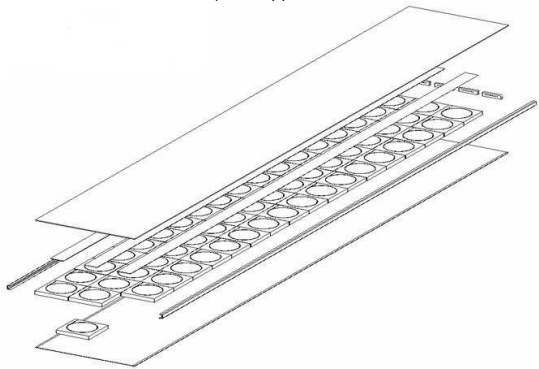,angle=0,width=17.2pc}}
\vspace*{-.38in}
\caption{View of the CPR upgrade design.}
\label{fig:CPR2}
\vspace*{-.4in}
\end{figure}

The new CPR will be based on 2cm thick scintillator tiles
segmented in $\eta$ and $\phi$ and read out by a 1mm diameter WLS fiber
running into a groove on the surface of each tile.
Six tiles (12.5x12.5 cm$^2$ each) will cover the front face of
each calorimeter tower, and the tiles will be assembled
in 48 modules like the one shown in Fig.~\ref{fig:CPR2} covering
the 48 central calorimeter wedges.
After leaving the tiles,
the WLS fibers will be spliced to clear fibers which
will terminate into plastic connectors at the
higher $\eta$ edge of each module. There $\sim$5m long optical cables will 
transmit the light to 16-channel
{\it PhotoMultiplier Tubes} (PMTs) at the back of the wedge.
A current prototype, consisting in scintillator tiles provided by the
CDF JINR Dubna group and Pol.Hi.Tech fibers, provided a light yield of
$\sim$20 (12) photoelectrons at the exit of the tile (after all the optical chain),
exceeding the design requirement.

The new CCR will use the same technique but the available space will limit the scintillator
thickness to 5mm. Ten tiles, $\sim$5cm wide, will cover each $\phi$-crack with the
same calorimeter segmentation of 10 towers/wedge.
 
\section{THE EM TIMING PROJECT}
\vspace*{-.05in}
The CDF Collaboration is adding timing information into the readout of
the Central and Plug electromagnetic calorimeters (CEM and PEM) using a technique
similar to the hadron TDC system.
This upgrade would significantly improve the potential of the CDF detector
to do high-$P_T$ searches for new physics in data samples with photons in the
final state by: 1) reducing the the cosmic ray and beam halo sources
of background; 2) checking that all photons in unsual events are from the
primary interaction. 
With sufficient calibration data, there is even the possibility of searching
for very long-lived particles which decay (1-10ns) into photons.

\vspace*{-.05in}
\subsection{The Hardware Project}
The signal from the PMT goes to a Transition Board on the back
of a calorimeter readout VME crate. All the lines are passed through
the VME backplane into an Amplifier Shaper Discriminator (ASD) which
effectively turns the signal into an LVDS digital pulse
suitable for use by a TDC.

While the 960 PEM PMTs already have a dynode output designed into the base, 
a custom {\it splitter} is used for the output
of the 960 CEM PMTs.
The splitter is a fully passive element, completely connectorized,
which works by inductively coupling the primary line (for the
energy measurement) to the secondary output.
The primary output loses a negligible amount of the charge and the
secondary line only takes $\sim$15\% of the output voltage for use to fire the
ASD/TDC system.

Several tests of the splitter, both on the test bench and on the detector itself,
show no difference between the input shape and height before and after
inserting the splitter into the system.
For energies $>$ 4 (2) GeV for the CEM (PEM), 
the splitter fires the ASD/TDC system with 100\% efficiency.
The intrinsic timing resolution, measured using the CEM LED
and a splitter into a TDC channel of the hadron calorimeter,
is $\sim$1.1ns,
dominated by the 1ns TDC resolution.


\end{document}